\documentclass [12pt]{article}
\usepackage[dvips]{graphicx}

\usepackage{epsfig}
\usepackage{graphicx}
\usepackage{indentfirst}
\usepackage{amsmath}
\usepackage{amsfonts}
\usepackage{amssymb}
\usepackage{hyperref}
 \usepackage{latexsym}
\usepackage{color}

\begin{document}
 \begin{center}
{\Large\bf$f(T)$ gravity and energy distribution in Landau-Lifshitz
prescription }\\
\medskip

 M. G. Ganiou $^{(a)}$\footnote{e-mail:moussiliou\_ganiou @yahoo.fr},  
 M. J. S. Houndjo$^{(a,b)}$\footnote{e-mail:sthoundjo@yahoo.fr} 
 and J. Tossa$^{(a)}$\footnote{e-mail: joel.tossa@imsp-uac.org}

 $^a$ \,{\it Institut de Math\'{e}matiques et de Sciences Physiques (IMSP)}\\
  {\it 01 BP 613,  Porto-Novo, B\'{e}nin}\\
 $^{b}$\,{Universit\'e de Natitingou - B\'enin} \\
\end{center}
\begin{abstract}
We  investigate in this paper the  Landau-Lifshitz energy distribution in the framework of $f(T)$
theory view as a modified version of Teleparallel theory. From some important Teleparallel theory results on the localization of energy,
our investigations  generalize the Landau-Lifshitz prescription  from the computation of the energy-momentum complex to the framework of $f (T)$ gravity  as it is done in the modified versions of General Relativity.
We compute the energy density  in the first step for three  plane symmetric metrics in vacuum. We find  
 for the second metric that the energy density vanishes independently of $f (T)$ models. These metrics provide results in perfect agreement with those
mentioned in literature. In the second step the calculations are performed for the  Cosmic String Spacetime metric. It results  that    
  the energy distribution depends on  the mass $M$ of cosmic string and it is strongly affected by the parameter of 
  the considered $f (T)$ quadratic model.
\end{abstract}
 
\section{Introduction}
To express  the energy-momentum as a unique tensor quantity or in other formulation,
the problem of the localization of energy in General Relativity, has been since an open problem in this branch of theoretical physic.
Indeed, the localization of energy-momentum  is one of the oldest and thorny problems \cite{Misner} in
General Relativity (GR) which is still without any acceptable answer in general. In order to solve the problem, several attempts have been made,
starting by Einstein who was the first to introduce  locally conserved energy-momentum pseudo tensors  
known as energy-momentum complex\cite{Trautman}.
Being a natural field, it is expected that gravity should have its own local energy-momentum density.  
Misner, Thorne and Wheeler\cite{Misner} argued that the energy is localizable
only for spherical systems. This viewpoint has been contradicted by Cooperstock and Sarracino \cite{Cooperstock}. 
They stipulated that if energy can be localized in spherical systems, it can also be localized for all systems.
Bondi \cite{ Bondi} agreed with them and argued  that one can not admit in GR   a non-localizable form of energy 
whose location can in principle be found. 
 Through several very interesting works, 
different definitions for the energy and momentum distributions have been introduced by others  physicists in  RG and others equivalent 
theories, namely M\O{}ller \cite{Møller},
 Tolman\cite{Tolman}, Landau-Lifshitz  \cite{Landau},  Papapetrou \cite{Papapetrou}, Bergmann-Thomson \cite{Bergmann}, 
 Weinberg\cite{Weinberg}, Qadir-Sharif\cite{Qadir}, Mikhail\cite{Mikhail}, Vargas\cite{Vargas}. 
 All these prescriptions, except M\O{}ller, are restricted to perform calculations in Cartesian coordinates only. Recently Virbhadra 
 \cite{Virbhadra0} remarked that the concept of energy-momentum complexes are very useful in investigating
 the Seifert conjecture for naked singularities and the hoop conjecture of
Thorne. By considering this fundamental result, Chang
and his collaborators \cite{Chang} proved that every energy-momentum complex can be associated
with a particular Hamiltonian boundary term. Thus the energy-momentum complexes may also be considered as quasi-local. 
Landau-Lifshitz \cite{Landau} introduced  the energy-momentum complex by using
the geodesic coordinate system at some particular point of space.
Why studying the gravitational energy density? The gravitational energy density plays a remarkable role in the description 
of the total energy of the universe. Firstly, it is mentioned in literature that at the time of the creation of our universe which results 
from quantum fluctuation of the vacuum, any conservation law of physics need not to have been violated \cite{Tryon}. 
Indeed, Tryon \cite{Tryon} in same idea as Albrow \cite{Albrow} 
suggested  that our universe must have a zero net value for all conserved quantities. They supported their  point of 
view by using a Newtonian order of magnitude estimate and obtained
 that the net energy of our universe may be indeed zero. 
 By using Killing vectors, Cooperstock and Israelites \cite{Israelit} have initiated a real and 
 interesting work on the energy-momentum distributions of the open and closed universes
They found  zero as the  value of energy for any homogeneous isotropic universe described by
a Friedmann-Robertson-Walker metric in the context of General Relativity. 
This result has been confirmed by many other authors through  different works on metrics describing these kinds of universes
\cite{Vargas,Rosen}. Secondly, in attempt to answer the previous question,
some authors like \cite{Prigogine} and it collaborators,
argued that  during inflation the vacuum energy driving the accelerated 
expansion of the universe, and which was responsible for the creation of radiation
and matter in the universe, is drawn from the energy of the gravitational field. \par 
Recently, the problem of energy-momentum localization was also considered in modified theories of General Relativity, 
the so called $f(R)$ theory  \cite{Putaja,Sharif} as in Teleparallel gravity \cite{Mikhail,Vargas,Nashed0} and its modified version namely 
$f(T)$ theory\cite{Ulhoa}. Teleparallel gravity is an alternative description of gravitation which corresponds to
a gauge theory for the translation group \cite{ Aldrovandi}. There are already no bad number of papers, 
on the energy-momentum localization in this theory. The authors found
that the problem of energy-momentum localization can also be discussed in this alternative
theory, their results are in agreement with those found in GR. For example, in \cite{Vargas},
Vargas established the Einstein and Landau-Lifshitz energy-momentum complex in the framework of Teleparallel gravity and 
found zero for   the total energy  in
Friedmann-Robertson-Walker space-times. Among other interesting works, we can note that of
Of Mustafa \cite{Mustafa}. The author used the M\O{}ller, Einstein, Bergmann-Thomson, and Landau-Lifshitz prescriptions
in both GR and Teleparallel gravity to evaluate the energy distribution for the generalized Bianchi-type I metric.
For the first time in literature, Ulhoa and Spaniol \cite{Ulhoa}  
presented and analyzed an expression for the gravitational energy-momentum vector in the context of $f(T)$ theories. 
 They also  obtained a vanishing gravitational energy for a particular choice of the tetrad.\\
  In this paper, we propose to  investigate the same aspect of Landau-Lifshitz  in the context of the modified Teleparallel $f(T)$ gravity.
  Our main goal in  this
work is  to extract the Landau-Lifshitz energy-momentum complex valid for $f(T)$ theory. Then, we aim 
to evaluate the corresponding energy density for some particular metrics. Our investigation was motivated by the fact that 
 the  Landau-Lifshitz   energy-momentum complex  in particular has been recently at the center
 of very interesting discussions in GR and in the framework of $f(R)$ theory.   
The Laudau-Lifshitz energy distribution has been   calculated for two metrics which describe
  non-asymptotically flat black hole solutions in dilaton-Maxwell gravity \cite{IRINA}. This investigation shows that the obtained
  energies depended on the mass, the charge of the black hole and the parameter of these metrics. In addition, the authors
in \cite{Putaja} generalized for the first time the Landau-Lifshitz prescription from  the computation
    of  the energy-momentum complex to the framework of $f (R)$ gravity with constant scalar curvature as necessary condition. Some time after 
 Sharif and  Shamir \cite{Sharif} used  this energy-momentum complex 
to evaluate  in the framework of $f(R)$, the energy density of  three plane symmetric metrics  and for cosmic string spacetime. \\
The paper is organized as follows. In section Sec.II, we review some  fundamental results on  the Landau-Lifshitz 
energy-momentum complex in the Teleparallel theory. This  have allowed us to establish at the end of the section,
the generalized Landau-Lifshitz 
energy-momentum complex  valid for  the so-called $f(T)$ gravity. In section Sec.III, 
we compute  the obtained energy density for  plane symmetric metrics. 
The section Sec.III is devoted to energy distribution of cosmic string spacetime metric. The conclusion 
comes in the last section. 

\section {Teleparallel and $f(T)$  versions of 
   Landau-Lifshitz  Energy-Momentum Complexe ($EMC$)}
Through  the revision of the fundamental results on the Landau-Lifshitz  Energy-Momentum Complex    in the Teleparallel theory,
we will establish
the  generalized Landau-Lifshitz   energy-momentum complex  valid for the $f(T)$ theory. 
Before giving the Teleparall version of the Landau-Lifshitz energy-momentum complex, we briefly
outline the main points of the Teleparall theory. In general,  
when formulating theories of gravity, the metric tensor is of paramount importance. 
It contains the information needed to locally measure distances and thus
to make theoretical predictions about experimental findings. Furthermore, the structure of the spacetime can be described by   an
alternative dynamical variable, the well known  non-trivial tetrad $h^a_{\;\;\mu}$ which is a set of four
vectors defining a local frame at every point. The tetrads represent the   basic entity of the theory of
Teleparallel gravity. From their reconstruction  arises the  Teleparall theory as gravitational theory naturally based on the
gauge approach of the group of translations. The tretrads are  defined from  the gauge covariant derivative
for a scalar field as $h^a_{\;\;\mu}= \partial_\mu x^a+A^a_{\;\;\mu}$ with $A^a_{\;\;\mu}$ the translational gauge potential and $x^a$
  the tangent-space coordinates \cite{de Andrade}. The tretrad $h^a_{\;\;\mu}$ and its inverse $h_a^{\;\;\mu}$ satisfy the following relations: 
\begin{equation}
  h^a_{\;\;\mu}h_a^{\;\;\nu}=\delta^\nu_\mu \quad \quad h^a_{\;\;\mu}h_b^{\;\;\mu}=\delta^a_b. 
\end{equation}
   An another important notion resulting from the establishing of this theory is
   { \it the condition of absolute parallelism} \cite{Hayashi} which leads to the Weitzenb\"{o}ck  connection  seen  as 
   the fundamental connection of the theory. It is 
     given by
 \begin{equation}
 \Gamma_{\mu\nu}^{\lambda}= h_a^{\;\;\lambda}\partial_\nu h^a_{\;\;\mu}
 =-h^a_{\;\;\mu}\partial_\nu h_a^{\;\;\lambda}.
 \end{equation}.  
 We emphasize here that    the
Latin alphabet $(a, b, c, ... = 0, 1, 2, 3)$ is used to denote the tangent space
indices and the Greek alphabet $(\mu,\nu, \rho, ... = 0, 1, 2, 3)$ to denote the spacetime
indices. The metric and the tetrad are related by 
\begin{equation}
  g_{\mu\nu}= \eta_{ab}h^a_{\;\;\mu}h^b_{\;\;\nu},
\end{equation}
where  $\eta_{ab}=\text{diag}(+1,-1,-1,-1)$ is the Minkowski metric of the tangent space. 
In Teleparallel gravity and due to the  no curvature  Weitzenb\"{o}ck  connection,  the effects of gravitation are described by the
torsion tensor, while curvature tensor  does not appear. Consequently, the non-vanishing  and naturally antisymmetric 
torsion  tensor is expressed via its components by  
\begin{equation}
   T_{\;\;\;\mu\nu}^{\lambda}= \Gamma_{\;\;\;\nu\mu}^{\lambda}-
   \Gamma_{\;\;\;\mu\nu}^{\lambda}=h_a^{\;\;\lambda}(\partial_\mu h^a_{\;\;\nu}-\partial_\nu h^a_{\;\;\mu})\neq 0.
\end{equation}

Another important tensor emerging from the use of the Weitzenb\"{o}ck connection 
is the contortion tensor $  K_{\;\;\;\mu\nu}^{\;\,\lambda}$ which shows the difference between 
the  Weitzenb\"{o}ck connection and the Levi-Civita  connection \cite{equiv} according to
\begin{center}
\begin{equation}\label{cont}
  K_{\;\;\;\mu\nu}^{\lambda}:= \Gamma_{\;\;\;\mu\nu}^{\lambda}-
  \tilde{\Gamma}_{\;\;\;\mu\nu}^{\lambda}=\frac{1}{2}\Big (T_{\nu}{}^{\lambda}{}_{\mu}+
  T_{\mu}{}^{\lambda}{}_{\nu}- T^{\lambda}{}_{\mu\nu}\Big),
\end{equation}
\end{center}
where $ \tilde{\Gamma}_{\;\;\;\mu\nu}^{\;\,\lambda}$  are the Christoffel symbols or the coefficient of Levi-Civita connection.
The action of Teleparallel gravity in the presence of matter is given by
  \begin{equation}\label{act1}
  S=\frac{1}{\kappa^2}\int d^4x hT+ \int d^4x h\mathcal{L}_M,
 \end{equation}
   where $h=|\text{det}(h^a{}_\mu)|$ is equivalent to 
 $\sqrt{-g}$ in  General Relativity, $\kappa^2=\frac{16\pi G}{c^4}$,  $\mathcal{L}_M$ is the Lagrangian of the matter field
 and  $T$ the scalar  torsion defined by  
 \begin{equation}\label{de}
  T:=S_\beta{}^{\mu\nu}
 T{}^\beta{}_{\mu\nu},
 \end{equation}
with  $S_\beta{}^{\mu\nu}$, specifically defined by \cite{Hayashi}
\begin{equation}\label{sup}
 S_\beta{}^{\mu\nu}=\frac{1}{2}\Big(  
 K{}^{\mu\nu}{}_{\beta}+
 \delta^\mu_\beta T{}^{\alpha\nu}{}_{\alpha}- \delta^\nu_\beta T{}^{\alpha\mu}{}_{\alpha}  \Big).
\end{equation}
The variation of  (\ref{act1})  with respect to $h^a{}_\mu$ leads to the Teleparallel
field equations with $c^4=1$,
\begin{equation}\label{fiel}
\partial_\nu(hS_\beta{}^{\mu\nu}) - 4\pi G(ht^\mu{}_\beta)= 4\pi Gh T^\mu{}_\beta, 
\end{equation}
where $T^\mu{}_\beta=-h^{-1}h^a_{\;\;\beta}\Bigg[\frac{\partial \mathcal{L}_m}{\partial h^a_{\;\;\mu}}-
\partial_\alpha \frac{\partial\mathcal{L}_m }{\partial(\partial_\alpha h^a_{\;\;\mu })}\Bigg]$  
is the energy-momentum tensor of  the matter fields and 
\begin{equation}\label{psedo}
 t^\mu{}_\beta= \frac{1}{4\pi G}h\Gamma_{\beta\lambda}^{\nu}S_\nu{}^{\mu\sigma}-\delta^\mu_\lambda \mathcal{L}_G
\end{equation}
is the energy-momentum pseudotensor of the gravitational field 
\cite{Guillen}. Furthermore,  the equation (\ref{fiel}) can be rewritten in the following relation 
\begin{equation}\label{fiel1}
h(t^\mu{}_\beta+ T^\mu{}_\beta)= \frac{1}{4\pi G} \partial_\nu(hS_\beta{}^{\mu\nu}).  
\end{equation}
 From this relation and due  to  the antisymmetry  of the tensor $S_\beta{}^{\mu\nu}$ in its last indices, one can extract
 the conservation law in the following relation
\begin{equation}\label{fiel1}
 \partial_\mu [h(t^\mu{}_\beta+ T^\mu{}_\beta)]=0.
\end{equation}
According to the equivalence between the Teleparallel equation and the Einstein's equations \cite{de Andrade}, the Teleparallel tensor
$U_\beta{}^{\mu\nu}=hS_\beta{}^{\mu\nu}$ stays  for the
Freud's superpotential  \cite{Vargas},\cite{Sezgin}, \cite{Mustafa} and \cite{Salti}. 
Consequently, $t^\mu{}_\beta$ is nothing but the Teleparallel version of
Einstein's gravitational energy-momentum pseudotensor.  
In addition this Teleparallel version of Freud's superpotential is directly related
 to the geometrical density Lagrangian $\mathcal{L}_G$ via the fundamental relation \cite{Vargas} 
\begin{equation} \label{super}
 U_\beta{}^{\mu\nu}= 4\pi G  h^a_{\;\;\beta} \frac{\partial\mathcal{L}_G }{\partial(\partial_\nu h^a_{\;\;\mu })} 
\end{equation}
The Landau-Lifshitz's energy-momentum complex
in Teleparallel gravity   given as follow \cite{Vargas}

 \begin{equation}\label{liftel}
  hL^{\mu\nu}=\frac{1}{4\pi G}\partial_\sigma(h g^{\mu\lambda}U_\beta{}^{\nu\sigma}).
 \end{equation}
  According to (\ref{fiel1}), the Landau-Lifshitz's energy-momentum complex $L^{\mu\nu}$ satisfies the local
conservation laws $\frac{\partial L^{\mu\nu}}{\partial x^\nu}=0$ . 

The energy and momentum distributions of Landau-Lifshitz prescription  in the Teleparallel gravity are summarized in  the
following equation \cite{Vargas}:
\begin{equation}\label{ener}
 P_\mu=\int_\sum hL^0{}_\mu dxdydz, 
\end{equation}
where $ P_\mu$ for $\mu = 1, 2, 3 $  gives the momentum components and $ P_0$  stays for the energy. The
 integration hypersurface $\sum$ is described by $x^0=t=$constant. In the following paragraph, we are searching for the expression the 
of Landau-Lifshitz's energy-momentum complex in the framework a  Teleparallel modified version namely $f(T)$ theory. \par
The action of the modified versions of  TEGR (Teleparallel equivalent of General Relativity \cite{So}  
is obtained by substituting the scalar torsion 
of the action (\ref{act1}) by an arbitrary function of scalar torsion   obtaining modified theory $f(T)$. 
This approach is similar in spirit to the generalization of 
Ricci scalar curvature of Einstein-Hilbert action  by a function of this scalar  leading  to the well known 
$ F(R) $ theory.  Indeed,  the action of $f(T)$ theory can be defined  as 
 \begin{equation}\label{act2}
  S=\frac{1}{\kappa^2}\int d^4x hf(T)+ \int d^4x h\mathcal{L}_M.
 \end{equation}
The variation of this action with respect to tetrad $h_a{}^{\mu}$ gives 
( \cite{Ulhoa}, \cite{Houndjo} and \cite{salako})
 \begin{eqnarray}\label{mot}
  \frac{1}{h}\partial_\mu(hS_a{}^{\mu\nu})f_T(T)- h_a{}^{\lambda}T^\rho{}_{\mu\lambda}S_\rho{}^{\mu\nu}f_T(T) 
  +S_a{}^{\mu\nu}\partial_\mu(T)f_{TT}(T)+\frac{1}{4h} h_A{}^{\nu}f(T)= \frac {1}{4\kappa^2}T^\nu_a, 
 \end{eqnarray}
 with $f_T(T)=df(T)/dT$,   $f_{TT}(T)=d^2f(T)/dT^2$.  The field equations can be recast in the following form  \cite{Ulhoa})
\begin{eqnarray}\label{lif0}
 \partial_\sigma \Big[h S^{a\tau\sigma}f_T(T)\Big]=\frac{1}{4\kappa^2}h^a{}_{\lambda}\Big(t^{\tau\lambda}+T^{\tau\lambda}\Big),
\end{eqnarray}
where
\begin{eqnarray}\label{pseudo}
 t^{\tau\lambda}=\frac{1}{\kappa^2}\Big[4f_T(T)g^{\lambda\beta}S_{\mu}{}^{\nu\tau}T^\mu{}_{\nu\beta}-g^{\tau\lambda} f(T)\Big]. 
\end{eqnarray}
Due to the skew-symmetry of $S_{\mu}{}^{\nu\tau}$ in the  two up indices, it follows that 
\begin{equation}
 \partial_\tau\partial_\sigma [h S^{a\tau\sigma}]\equiv 0,
\end{equation}
and one obtains 
\begin{equation}
 \partial_\tau(ht^{a\tau}+hT^{a\tau})=0. 
\end{equation}
We note here  that the above expression works as a conservation law for the sum
of the energy-momentum tensor of matter fields and the quantity $t^{\tau\lambda}$ . Thus $t^{\tau\lambda}$ is
interpreted as the energy-momentum tensor of the gravitational field in the context
of $f (T)$ theories \cite{Ulhoa}, being more general than (and slightly different from) the usual
quantity in Teleparallel presented in (\ref{psedo}).
However, making using the Laudau-Lifshitz   prescription  \cite{Landau}  and its applications in the context of Teleparallel gravity as it
was done in \cite{Vargas}, one can rewrite the equation (\ref{lif0}) in the following form
\begin{equation}
 \frac{1}{4\pi G}\partial_\nu [g^{\tau\sigma}h \tilde{U}_\sigma{}^{\lambda\nu}]=h^2(t^{\tau\lambda}+T^{\tau\lambda}),
\end{equation}
where 
\begin{equation}\label{fun}
 \tilde{U}_\sigma{}^{\lambda\nu}=4\pi G  h^a_{\;\;\sigma} \frac{\partial\tilde{\mathcal{L}}_G }{\partial(\partial_\lambda h^a_{\;\;\nu})},
\end{equation}
with $\tilde{\mathcal{L}}_G =\frac{1}{\kappa^2}hf(T)$. Consequently, the tensor $\tilde{U}_\sigma{}^{\lambda\nu}$ can be interpreted as the 
Freud's super-potential tensor in the contexte of $f (T)$ theory  and it generalizes  its Teleparallel version found by  \cite{Vargas} (i.e
for $f(T)=T$ we get $U_\sigma{}^{\lambda\nu}$ in (\ref{super})). In addition, Following the same approach as \cite{Vargas}, we  deduct simply
the $f(T)$ version of the Laudau-Lifshitz energy-momentum complex as 
 \begin{equation}\label{dor}
  h\tilde{L}^{\tau\lambda}=\frac{1}{4\pi G}\partial_\nu [g^{\tau\sigma}h \tilde{U}_\sigma{}^{\lambda\nu}],
 \end{equation}
or in equivalent form as it is established in General Relativity \cite{Mustafa,Salti}  
and its modified versions \cite{Putaja,Sharif } as
\begin{equation}
  h\tilde{L}^{\tau\lambda}=h^2(t^{\tau\lambda}+T^{\tau\lambda}).
\end{equation}
Here $h^2$ can be identified to $-g$ in General Relativity. As conclusion,  $T^{\tau\lambda}$ 
is usually  the energy-momentum tensor of the matter and all non-gravitational fields,
and $t^{\tau\lambda}$ in (\ref{pseudo}) is the $f(T)$  version of Landau-Lifshitz energy-momentum pseudo tensor. So, the locally
conserved quantity $h\tilde{L}^{\tau\lambda}$ contains contributions from the matter, non-gravitational fields
and gravitational fields. \par
In order to facilitate the calculation, we rewrite the generalized Landau-Lifshitz energy-momentum complex
 $h\tilde{L}^{\tau\lambda}$ of (\ref{dor}) in term  of Teleparallel quantities 
via the following relation,
\begin{equation}
  h\tilde{L}^{\tau\lambda}= f_T(T)hL^{\tau\lambda}+\frac{1}{4\pi G}hg^{\tau\sigma}U_\sigma{}^{\lambda\nu} f_{TT}(T)\partial_\nu T,
\end{equation}
where $hL^{\tau\lambda}$  is the Landau-Lifshitz  energy-momentum complex evaluated in the framework of Teleparallel gravity ( see (\ref{liftel}))
and $U_\sigma{}^{\lambda\nu}$, the  corresponding Freud's super-potential. A  such expression is more general namely it
is valid for any  algebraic function of scalar torsion $f(T)$ contrarily to the case of $f(R)$ gravity 
where the generalized Landau-Lifshitz energy-momentum is 
only valid for constant curvature models \cite{Putaja,Sharif}. The
completely timelike component of $h\tilde{L}^{\tau\lambda}$ had the mathematical features of an energy density. 
Then, we extract the generalized energy density from  the above equation as 
\begin{equation}\label{bigre}
 h\tilde{L}^{00}= f_T(T)hL^{00}+\frac{1}{4\pi G}hg^{0\sigma}U_\sigma{}^{0\nu} f_{TT}(T)\partial_\nu T.
\end{equation}
One can also have 
\begin{equation}\label{bigre1}
  h\tilde{L}^{00}=h^2(t^{00}+T^{00}).
\end{equation}
It would be worthwhile to mention here that we need cartesian coordinates to use these formulas. Indeed, it was shown in \cite{Virbhadra} 
that the different energy-momentum complexes restricted to
cartesian coordinates, give the same and acceptable energy distribution for a given spacetime. A parallel analysis was done by \cite{Vargas},
with the conclusion that it is important to perform in  cartesian coordinates, as any other coordinate may lead to
non-physical values for the pseudotensor. 

\section{Energy distribution  of plane symmetric solutions}
 We  evaluate energy density of three  plane symmetric  solutions  from \cite{Bertolami}  in order to apply  the  found $f(T)$ version
of the Laudau-Lifshitz energy-momentum  complex. 
\subsection{Application to  Taub's metric: first solution }
 The energy density of the first vacuum solution that we explore here, concerns the Taub’s metric given by 
\begin{equation}\label{met1}
 ds^2=k_1x^{\frac{-2}{3}}dt^2-dx^2-k_2x^{\frac{4}{3}}(dy^2+dz^2),
\end{equation}
where $k_1$ and $k_2$ are constants. Concerning the vierbein choice
that realizes the above metric we choose the diagonal one,
namely
\begin{eqnarray}\label{tetr1}
 h^a{}_{\mu}=\text{diag}(k_1^{\frac{1}{2}}x^{\frac{-1}{3}}, 1,k_2^{\frac{1}{2}}x^{\frac{2}{3}},k_2^{\frac{1}{2}}x^{\frac{2}{3}})\quad
  h_a{}^{\mu}=\text{diag}(k_1^{-\frac{1}{2}}x^{\frac{1}{3}}, 1,k_2^{-\frac{1}{2}}x^{\frac{-2}{3}},k_2^{\frac{-1}{2}}x^{\frac{-2}{3}}).
 \end{eqnarray} 
   One has $h=k_1^{\frac{1}{2}}k_2x$. From Eqs.(\ref{met1}) and (\ref{tetr1}), and by adopting the notation $(t,x,y,z)=(x^0,x^1,x^2,x^3)$,
we can now construct the Weitzenb\"{o}ck connection, whose nonvanishing components are found in
\begin{equation*}
 \Gamma^0{}_{01}=-\frac{1}{3x}, \quad   \Gamma^2{}_{21}=\Gamma^3{}_{31}=\frac{2}{3x}.
\end{equation*}
The corresponding  non-vanishing components of torsion  tensor are
\begin{eqnarray*}  
 T^0{}_{01}=-T^0{}_{10}=\frac{1}{3x}\quad \text{and} \quad
  T^2{}_{12}=T^3{}_{13}=-T^2{}_{21}=-T^3{}_{31}= \frac{2}{3x}.
\end{eqnarray*}
Now, the non-zero components of the tensor $S_\sigma{}^{\mu\nu}$ read
\begin{eqnarray*}
 S_0{}^{10}=-S_0{}^{01}=\frac{2}{3x}   \quad \text{and} \quad
  S_2{}^{12}=S_3{}^{13}=-S_2{}^{21}=-S_3{}^{31}= \frac{1}{6x}.
\end{eqnarray*}
The scalar torsion also reads $T=0$. The Teleparallel Freud tensor components read
\begin{eqnarray*}
 U_0{}^{10}=-U_0{}^{01}=\frac{2}{3}k_1^{\frac{1}{2}}k_2  \quad \text{and} \quad
 U_2{}^{12}=U_3{}^{13}=-U_2{}^{21}=-U_3{}^{31}= \frac{1}{6}k_1^{\frac{1}{2}}k_2. 
\end{eqnarray*}
The $hL^{00}$  becomes
\begin{equation}
 hL^{00}=-\frac{5x^{\frac{-2}{3}}}{18\pi G}k_2^2,
\end{equation}
whereas $h\tilde{L}^{00}$, according to null scalar torsion of the model, is expressed   by   
\begin{equation}
 h\tilde{L}^{00}= -\frac{5x^{\frac{-2}{3}}}{18\pi G}k_2^2f_T(T).
\end{equation}
It results that for all gravitational lagrangian densities of the form $f(T)=T+g(T)$ such as $g_T(0)=0$,  the 00-component of the
generalized Landau-Lifshitz is the same as in Teleparallel theory  for Taub's metric.
 
\subsection{Application to the second vacuum solution}
 The second vacuum solution is
\begin{equation}
 ds^2= (bx+bc)^2dt^2-dx^2-e^{2a}(dy^2+dz^2),
\end{equation}
where $a$,$b$ and $c$ are constants. We  build  the corresponding  tetrad components and its inverse  as 
\begin{eqnarray}\label{tetr2}
 h^a{}_{\mu}=\text{diag}((bx+bc), 1,e^{a},e^{a}),\quad
  h_a{}^{\mu}=\text{diag}((bx+bc)^{-1}, 1,e^{-a},e^{-a}),
 \end{eqnarray} 
with  $h=e^{2a}b(x+c)$. The nonvanishing components of the  torsion tensor  read
\begin{eqnarray}\label{tet1}
 T^0{}_{10}=&T^0{}_{01}=\frac{1}{x+c}. 
\end{eqnarray}
All the components of the  following Teleparallel tensors vanish: the tensor $S_\sigma{}^{\mu\nu}$, the  Freud's super-potential tensor. 
It follows $hL^{00}=h\tilde{L}^{00}=0$. The energy $P_0$  and the momentum components ($P_1$, $P_2$, $P_3$) also vanish and 
  agree with  some  interesting recent analysis made by \cite{Vargas} and \cite{Mustafa}. This also confirms the assumption of 
  others authors  \cite{ Tryon, Albrow},  when they assumed that the net energy of the universe may be equal to zero. 
  $P_0=0$ agrees with  Cooperstock and Israelit \cite{Israelit} results.

\subsection{Application to the third vacuum solution}
 The third solution  corresponds to anti de Sitter metric in GR and it is given
by
\begin{equation}\label{met3}
 ds^2= e^{2(c_1x+c_2)}(dt^2-dy^2-dz^2)-dx^2,
\end{equation}
with $c_1$ and $c_2$ constants. The  tretrads and its inverse  read 
\begin{eqnarray}\label{tetr3}
 h^a{}_{\mu}=\text{diag}\Big[e^{(c_1x+c_2)}, 1,e^{(c_1x+c_2)},e^{(c_1x+c_2)}\Big], \quad
  h_a{}^{\mu}=\text{diag}\Big[e^{-(c_1x+c_2)}, 1,e^{-(c_1x+c_2)},e^{-(c_1x+c_2)}\Big],
 \end{eqnarray} 
 with $h=e^{3(c_1x+c_2)}$.  The nonvanishing components of torsion tensor read
  \begin{eqnarray}\label{te21}
 T^0{}_{10}=-T^0{}_{01}=c_1, \quad  T^2{}_{12}= T^3{}_{13}=-T^2{}_{21}=-T^3{}_{31}=c_1.
\end{eqnarray}
   the non-zero components of the tensor $S_\sigma{}^{\mu\nu}$ read
  \begin{eqnarray*}
 S_0{}^{10}=-S_0{}^{01}=c_1  \quad \text{and} \quad
  S_2{}^{12}=S_3{}^{13}=-S_2{}^{21}=-S_3{}^{31}= c_1.
\end{eqnarray*}
The scalar torsion reads $T=6c^2_1$ and it is constant. The Teleparallel energy density  for this metric
is 
\begin{equation}
 hL^{00}=-\frac{c^2_1}{\pi G}e^{(c_1x+c_2)},
\end{equation}
whereas the generalized energy density becomes 
\begin{equation}\label{cdp}
 h\tilde{L}^{00}= -\frac{c^2_1}{\pi G}e^{(c_1x+c_2)}f_T(T).
\end{equation}
Considering  an important cosmological  Born-Infeld  model  \cite{Ferraro} 
\begin{equation}\label{ali}
 f(T)= T+ \lambda\Bigg[ \Big(1-\epsilon+\frac{2T}{\lambda}\Big)^{1/2}-1\Bigg],
\end{equation}
which satisfies the weak energy condition \cite{Di} for $\lambda>12.36$ and $0<\epsilon<1$. Taking into consideration the constant scalar 
torsion $T=6c^2_1$, the model becomes 
\begin{equation}
 f(T)=6c^2_1+ \lambda\Bigg[ \Big(1-\epsilon+\frac{12c^2_1}{\lambda}\Big)^{1/2}-1\Bigg]. 
\end{equation}
By making using of this value in (\ref{cdp}), one obtains 
\begin{equation}\label{cdpa}
 h\tilde{L}^{00}= -\frac{c^2_1}{\pi G}e^{(c_1x+c_2)}\Bigg[ 1+\Big(1-\epsilon+\frac{12c^2_1}{\lambda}\Big)^{-1/2}\Bigg].
\end{equation}
The model (\ref{ali}) must satisfy the relation $-3c^2_1f_T(T)+\frac{1}{4}f(T)=0$. This relation results from the application 
of the metric (\ref{met3}) to the motion equation (\ref{mot}) in vacuum. 
This equation has let us to constrain the constant $c_1$ by $\lambda$ through the following
solvable equation 
\begin{equation}
 \Big(\frac{3}{2}c^2_1+\frac{\lambda}{4}\Big)\Big(1-\epsilon+\frac{12c^2_1}{\lambda}\Big)^{1/2}=\frac{\lambda}{4}(1-\epsilon).
\end{equation}
   \section{  Energy Distribution of Cosmic String Space-time}
   The idea of big bang suggests that universe has expanded from a hot and
dense initial condition at some finite time in the past. It is a general cosmological 
assumption that the universe has gone through a number of phase
transitions at early stages of its evolution. During the expansion of the
 universe, cosmic strings would form a  cosmic network of macroscopic, quasi-stable
strings network that steadily unravels but survives to the present day, losing energy primarily by gravitational radiation \cite{Matt}.
Their gravity could have been responsible for the original clumping of matter into galactic superclusters. Cosmic strings \cite{David}, 
  if they exist, would be extremely thin with diameters on the same order as a proton. 
  They would have immense density, however, and so would represent significant gravitational sources. 
  A cosmic string $1.6$ kilometers in length may be heavier than the Earth. 
  However GR predicts that the gravitational potential of a straight string vanishes:
  there is no gravitational force on static surrounding matter. The only gravitational effect 
  of a straight cosmic string is a relative deflection of matter (or light) passing the string on opposite 
  sides (a purely topological effect). A closed loop of cosmic string gravitates in a more conventional way. Cosmic strings are one of the most
remarkable defects which are linear and string like. They have important
implications on cosmology such as large scale structures or galaxy formation. Therefore the nonstatic line element of the cosmic string has
 already been implicated in $f(R)$ gravity 
\cite{Sharif} to evaluate the energy distribution. Here, we will investigate
the contribution of this model in the context of energy distribution in the framework  of  $f(T)$ theory. To do so,
we consider the following non-static cosmic string space-time \cite{Abbassi}

\begin{equation}\label{met4}
 ds^2=dt^2-e^{2\sqrt{\frac{\Lambda}{3}t}}\Big[d\rho^2+(1-4GM)^2\rho^2d\phi^2+dz^2\Big], 
\end{equation}
 where $a=1-4GM=$, $\alpha=2\sqrt{\frac{\Lambda}{3}}$ with $G$,$M$ and $\Lambda$ are respectively the gravitational constant,
 mass per unit length of the string in the $z$ direction and the cosmological constant
.The  Laudau-Lifshitz prescription requires to work in Cartesian coordinates and consequently we rewrite the previous metric in terms of
Cartesian coordinates as 
\begin{equation}
  ds^2=dt^2-e^{\alpha t}\frac{x^2+a^2 y^2}{x^2+y^2}dx^2 -e^{\alpha t}\frac{y^2+a^2 x^2}{x^2+y^2}dy^2
  +2e^{\alpha t}xy\frac{a^2-1}{x^2+y^2}dxdy -e^{\alpha t}dz^2. 
\end{equation}
This metric can be constructed by  the  tetrad fields whose nonvanishing components and its inverse can be put in the following form

$$
h^a{}_\mu=\left(
\begin{array}{cccc}
1 & 0&0&0 \\
0 & \frac{x}{\sqrt{x^2+y^2}}e^{\beta t} &\frac{y}{\sqrt{x^2+y^2}}e^{\beta t}& 0 \\
\cr 0& -\frac{y}{a\sqrt{x^2+y^2}}e^{\beta t}& \frac{x}{a\sqrt{x^2+y^2}}e^{\beta t}&0\\
0&0&0&e^{\beta t}
\end{array}
\right),\quad 
h_a{}^\mu=\left(
\begin{array}{cccc}
1 & 0&0&0 \\
0 & \frac{x}{\sqrt{x^2+y^2}}e^{-\beta t} &-\frac{y}{a\sqrt{x^2+y^2}}e^{\beta t}& 0 \\
\cr 0& \frac{y}{a\sqrt{x^2+y^2}}e^{-\beta t}& \frac{x}{a\sqrt{x^2+y^2}}e^{-\beta t}&0\\
0&0&0&e^{-\beta t}
\end{array}
\right)
$$
The tetrads determinant also reads  $h=a e^{3\beta t}$. For these previous equations we have made $\alpha=2\beta$. Furthermore, 
the corresponding energy-momentum is given by  \cite{Sharif} 
\begin{equation}
 T^\mu_\nu=M\delta(x)\delta(y)\text{diag}(1,0,0,1).
\end{equation}
The nonvanishing components of the torsion  tensor read
 \begin{eqnarray}\label{refu}
 T^1{}_{10}=T^2{}_{20}=T^3{}_{30}=-\beta, \quad T^1{}_{21}= -T^1{}_{12}=\frac{y}{x^2+y^2},  \quad      
 T^2{}_{12}= -T^2{}_{21}=\frac{x}{x^2+y^2}.      
\end{eqnarray}
The nonvanishing components of the Teleparallel Freud tensor $U_\beta{}^{\mu\nu}=hS_\beta{}^{\mu\nu}$ are 
\begin{eqnarray*}
 U_0{}^{01}=-U_0{}^{10}&=&\frac{xy^2(a^2-1)e^{\beta t}}{2a(x^2+y^2)},\quad 
U_0{}^{02}=-U_0{}^{20}=\frac{y(x^2+a^2y^2)e^{\beta t}}{2a(x^2+y^2)},\quad
  U_3{}^{30}=-U_3{}^{03}= \frac{\alpha a}{8}(5e^{3\beta t}+e^{\beta t}),\\
 \cr  U_1{}^{10}=-U_1{}^{01}&=&\Big[\frac{5\alpha a}{8}+\frac{\alpha}{8a(x^2+y^2)^2}
 \Big(  a^2x^4+\alpha^2y^4+a^4x^2y^2+a^2x^2y^2  \Big)\Big]e^{3\beta t},\\
  \cr U_2{}^{20}=-U_2{}^{02}&=& \frac{5\alpha ae^{3\beta t}}{8}+\frac{ \alpha e^{3\beta t}}{8a(x^2+y^2)^2}\Big[(a^2x^2+y^2)(x^2+a^2y^2)-
 x^2y^2(a^2-1)^2  \Big],\\
   U_1{}^{21}=-U_1{}^{12}&=& \frac{y(x^2+a^2y^2)e^{\beta t}}{4a(x^2+y^2)^2}+  
 \frac{3x^2y(a^2-1)e^{\beta t}}{4a(x^2+y^2)^2}+\frac{e^{\beta t}}{4a^3(x^2+y^2)^4}[x^2y^2(a^2-1)^2-\\
 &&-(a^2x^2+y^2)(x^2+a^2y^2)],\\
  \cr U_1{}^{21}=-U_1{}^{12}&=& \frac{y(x^2+a^2y^2)e^{\beta t}}{4a(x^2+y^2)^2}+  
 \frac{3x^2y(a^2-1)e^{\beta t}}{4a(x^2+y^2)^2}+\frac{e^{\beta t}}{4a^3(x^2+y^2)^4}[x^2y^2(a^2-1)^2-\\
 &&-(a^2x^2+y^2)(x^2+a^2y^2)],\\
\cr  U_2{}^{12}=-U_2{}^{21}&=&\frac{e^{\beta t}}{4a(x^2+y^2)^2}[-xy^2(a^2-1)+x(a^2x^2+y^2)]
 +\frac{e^{\beta t}}{4a^3(x^2+y^2)^4}\Big[x^2y^2(a^2-1)^2-\\
 &&-(a^2x^2+y^2)(x^2+a^2y^2)\Big]\Big[xy^2(a^2-1)^2+x (a^2x^2+y^2)\Big],
 \cr U_3{}^{31}=-U_3{}^{13}&=&\frac{y^2x(a^2-1)e^{\beta t}}{2a(x^2+y^2)^2}, \quad
  U_3{}^{23}=-U_3{}^{32}=\frac{e^{\beta t}}{2a(x^2+y^2)^2}[(x^2+a^2y^2)+x^2y(a^2-1) ].
\end{eqnarray*}
The Teleparallel energy density according to the metric is also expressed as:
\begin{equation}\label{artis}
 hL^{00}=\frac{2e^{2\alpha t}}{\kappa^(x^2+y^2)^3}\Big[y^2(a^2-1)(y^2-3x^2)-x(x^2+3a^2y^2)(x^2+y^2)+4y^2(x^2+a^2y^2)\Big].
\end{equation}
In order to obtain the generalized Laudau-Lifshitz  energy density, we follow the same approach as \cite{Sharif}. This method allows us to 
express directly  the generalized Laudau-Lifshitz  energy density from its expression in (\ref{bigre1}). To do so, we calculate the 
00-component of the Laudau-Lifshitz energy-momentum tensor (\ref{pseudo}) as 
\begin{eqnarray}\label{prom}
t^{00}&=& \frac{1}{\kappa^2}f_T(T)\Bigg[ -\frac{15\alpha^2}{4}-\frac{\alpha^2}{4a^2(x^2+y^2)^2}\Big[ a^2(x^4+y^4)+  a^4x^2y^2
+a^2x^2y^2-\alpha^2 e^{-\alpha t} +    \nonumber \\
   && +(x^2+a^2y^2)(y^2+a^2x^2) -x^2y^2 (a^2-1)^2 \Big]   -\frac{f(T)}{f_T(T)} \Bigg]. 
\end{eqnarray}
We also evaluate $T^{00}$ as 
\begin{equation}
 T^{00}=M\delta(x)\delta(y). 
\end{equation}
We put all these expressions in (\ref{bigre1}), 
\begin{eqnarray}\label{lafai}
  h\tilde{L}^{00}&=& \frac{\alpha^2}{\kappa^2} e^{3\alpha t} f_T(T)\Bigg[ -\frac{15\alpha^2}{4}-\frac{\alpha^2}{4a^2(x^2+y^2)^2}\Big[ a^2(x^4+y^4)+  a^4x^2y^2
+a^2x^2y^2-\alpha^2 e^{-\alpha t} +  \nonumber\\ 
   && +(x^2+a^2y^2)(y^2+a^2x^2) -x^2y^2 (a^2-1)^2 \Big]   -\frac{f(T)}{f_T(T)} \Bigg]+ 
   a^2 e^{3\alpha t}M\delta(x)\delta(y) .
\end{eqnarray}
Before beginning discussing this generalized energy density for particular $f(T)$ model, 
let's present here the scalar torsion corresponding to the Cosmic String Space-time. 
\begin{eqnarray}
T&=&   -\frac{15\alpha^2}{8}-\frac{\alpha^2e^{-\alpha t} }{8}- \frac{\alpha^2}{8a^2(x^2+y^2)^2}\Big[ a^2(x^4+y^4)+ 
a^4x^2y^2+a^2x^2y^2 - \nonumber \\
&& -x^2y^2 (a^2-1)^2 +(x^2+a^2y^2)(y^2+a^2x^2)\Big]+\frac{xe^{-\alpha t}}{2a^2(x^2+y^2)^3}
\Big[x(a^2x^2+y^2)- \nonumber \\
&&-xy^2(a^2-1)\Big] -\frac{xe^{-\alpha t}}{2a^4(x^2+y^2)^5}
\Big[x^2y^2 (a^2-1)^2-(x^2+a^2y^2)(y^2+a^2x^2)     \Big]\times \nonumber \\
&&\times \Big[-xy^2 (a^2-1)^2-x (y^2+a^2x^2)    \Big] +\frac{y^2e^{-\alpha t}(x^2+a^2y^2)}{2a^2(x^2+y^2)^3}+
\frac{3y^2x^2(a^2-1)e^{-\alpha t}}{2a^2(x^2+y^2)^3}  + \nonumber \\
&&+\frac{ye^{-\alpha t}}{2a^4(x^2+y^2)^5}\Big[ x^2y^2 (a^2-1)^2-   (y^2+a^2x^2)  (x^2+a^2y^2)   \Big].
\end{eqnarray}
Let us remark here that the  scalar torsion associated to cosmic string space-time is not constant contrarily  
to its  scalar curvature \cite{Sharif}. \\
Considering the following important model  $f(T)$ \cite{Ulhoa}
\begin{equation}
 f(T)=T+\frac{1}{2}\lambda T^2.
\end{equation}
Such a quadratic model has been considered in several
cosmological contexts including inflation with  the graceful exit \cite{Nashed,Bamba} and mass of neutron stars in the presence 
of strong magnetic field \cite{Tossa}.  
By inserting it in the relation (\ref{lafai}), one has  

\begin{eqnarray}
  h\tilde{L}^{00}&=& 
\frac{1}{128 a^2 k \left(x^2+y^2\right)^6}e^{\alpha t} 
\Bigg[-16 y^2 \left(1+x^2 y\right)^2 \lambda +a^8 x^4 y^4 \left(-16+e^{2 \alpha t} \alpha^4 \left(x^2+y^2\right)^2\right) \lambda + \nonumber\\
 \cr &&+8 a^2 y \left(1+x^2 y\right) 
\Big(-\left(-4 \left(3 x^2 y^2+y^4\right)+\alpha^2 \left(x^2+y^2\right) \left(1+\left(x^2+y^2\right)^2\right)\right) \lambda +\nonumber\\
&&+e^{\alpha t} \left(x^2+y^2\right)^3
\left(8+15 \alpha^2 \lambda \right)\Big)+ \nonumber\\
\cr&&+2 a^6 x^2 y^2 \Big(e^{2 \alpha t} \alpha^4 \left(x^2+y^2\right)^2 \left(32 x^4+63 x^2 y^2+32 y^4\right) \lambda -
4 \Big(-4 \left(3 x^2 y^2+y^4\right)+\nonumber\\
\cr&&+\alpha^2 \left(x^2+y^2\right) \left(1+\left(x^2+y^2\right)^2\right)\Big) \lambda +e^{\alpha t} 
\left(x^2+y^2\right)^2 \left(32 \left(x^2+y^2\right)+\alpha^2 \left(-\alpha^2+
60 \left(x^2+y^2\right)\right) \lambda \right)\Big)+\nonumber\\
\cr&&+a^4 \Big[-\Big(-8 \alpha^2 y^2 \left(x^2+y^2\right) \left(3 x^2+y^2\right) \left(1+\left(x^2+y^2\right)^2\right)
+\alpha^4 \left(x^2+y^2\right)^4 \left(2+\left(x^2+y^2\right)^2\right)+\nonumber\\
\cr&&+16 y^3 \left(11 x^4 y+y^5+x^2 \left(2+6 y^3\right)\right)\Big) \lambda +
e^{2 \alpha t} \alpha^2 \left(x^2+y^2\right)^2 \times \nonumber\\
\cr&&\times \left(-240 \left(x^2+y^2\right)^4+\alpha^2 \left(17 x^4+33 x^2 y^2+17 y^4\right) \left(47 x^4+93 x^2 y^2+47 y^4\right) \lambda \right)
2 e^{\alpha t} \left(x^2+y^2\right)^2\times \nonumber\\
\cr&&\times \Big(-32 y^2 \left(x^2+y^2\right) \left(3 x^2+y^2\right)+
\alpha^4 \Big(x^4 \left(-17+15 x^4\right)+3 x^2 \left(-11+20 x^4\right) y^2+\left(-17+90 x^4\right) y^4+ \nonumber\\
\cr&&+60 x^2 y^6+15 y^8\Big)4 \alpha^2 \lambda\left(x^2+y^2\right) \left(2 \left(x^2+y^2\right) \left(1+\left(x^2+y^2\right)^2\right)-15 y^2 \left(3 x^2+y^2\right)
\lambda \right)\Big)\Big]\Bigg]+\nonumber\\
&&+a^2 e^{3\alpha t}M\delta(x)\delta(y).
\end{eqnarray}
By making using the approach followed by \cite{Vargas,Rosen,IRINA}, 
we obtain  from (\ref{ener}) the total energy  per unit length in the z direction as 

\begin{eqnarray}
E &=&\frac{1}{256 k (1-4 M)^2 r^8}e^{\alpha} \pi  
\Bigg[2 \Big(2+2 \left(-1+8 \left(M-8 M^3+8 M^4\right)\right) (\alpha-4 M \alpha)^2 r^4+ 
+\Big[(6+32 M \Big(-3+\nonumber\\
&& +2 M \left(7+8 M^2 \left(-7+8 M^2 (7+6 (-2+M) M)\right)\right)\Big)\Big] r^6-
(1-4 M)^4 \alpha^4 r^6 \left(-2+r^4\right)\Big) \lambda +\nonumber\\
&&+e^{2 \alpha t} (1-4 M)^4 r^{10} \left(256 k M-480 \alpha^2+\left(1598+M (-1+2 M) \left(128-3 M+6 M^2\right)\right) \alpha^4 \lambda \right)+\nonumber\\
&&+4 e^{\alpha} (1-4 M)^4 \alpha^2 r^6 \left(8 \left(-1+r^4\right)+\alpha^2 \left(17+M (-1+2 M)+15 r^4\right) \lambda \right)+\nonumber\\
&&+16 (1-4 M)^2 \left(-1+8 \left(M-8 M^3+8 M^4\right)\right) r^8 \left(-\alpha^2 \lambda +e^{\alpha t} \left(8+15 \alpha^2 \lambda \right)\right)
\text{Log}[r]\Bigg],
\end{eqnarray}
where  $r=\rho$.
Following the same approach as in \cite{IRINA}, we plot this energy  $E$ of the cosmic string as function of  its radius $r$ and  
unitary  mass $M$  and the correction parameter $\lambda$. Indeed by taking the cosmological 
constant as $\Lambda=1.3628\times10^{-60}\text{m}^{-2}$ (see \cite{ Nottale}), we
obtain the following figures.
\begin{figure}[h]
\centering
\includegraphics[width=8cm, height=8cm]{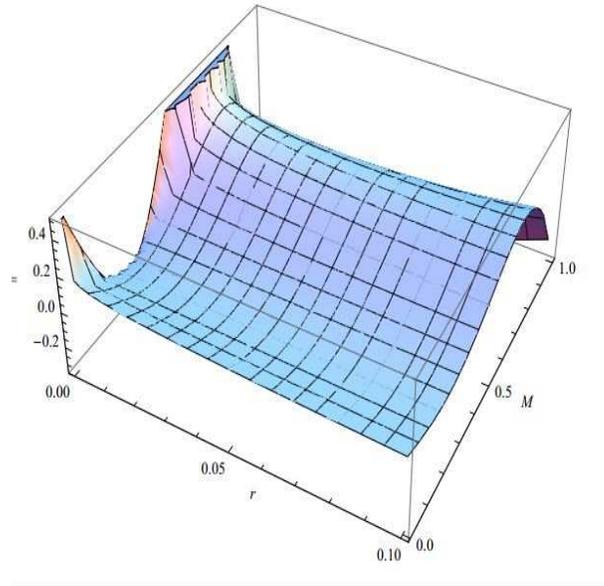}
  \caption{ The figure shows the variation of the total energy $E$ with respect to the radius 
  $r$ and the mass $M$ of the cosmic string for $\lambda$ with 
  corresponds to the case of Teleparallel.}
\label{fig1}
\end{figure}

\begin{figure}[h]
\centering
\begin{tabular}{rl}
\includegraphics[width=8cm, height=8cm]{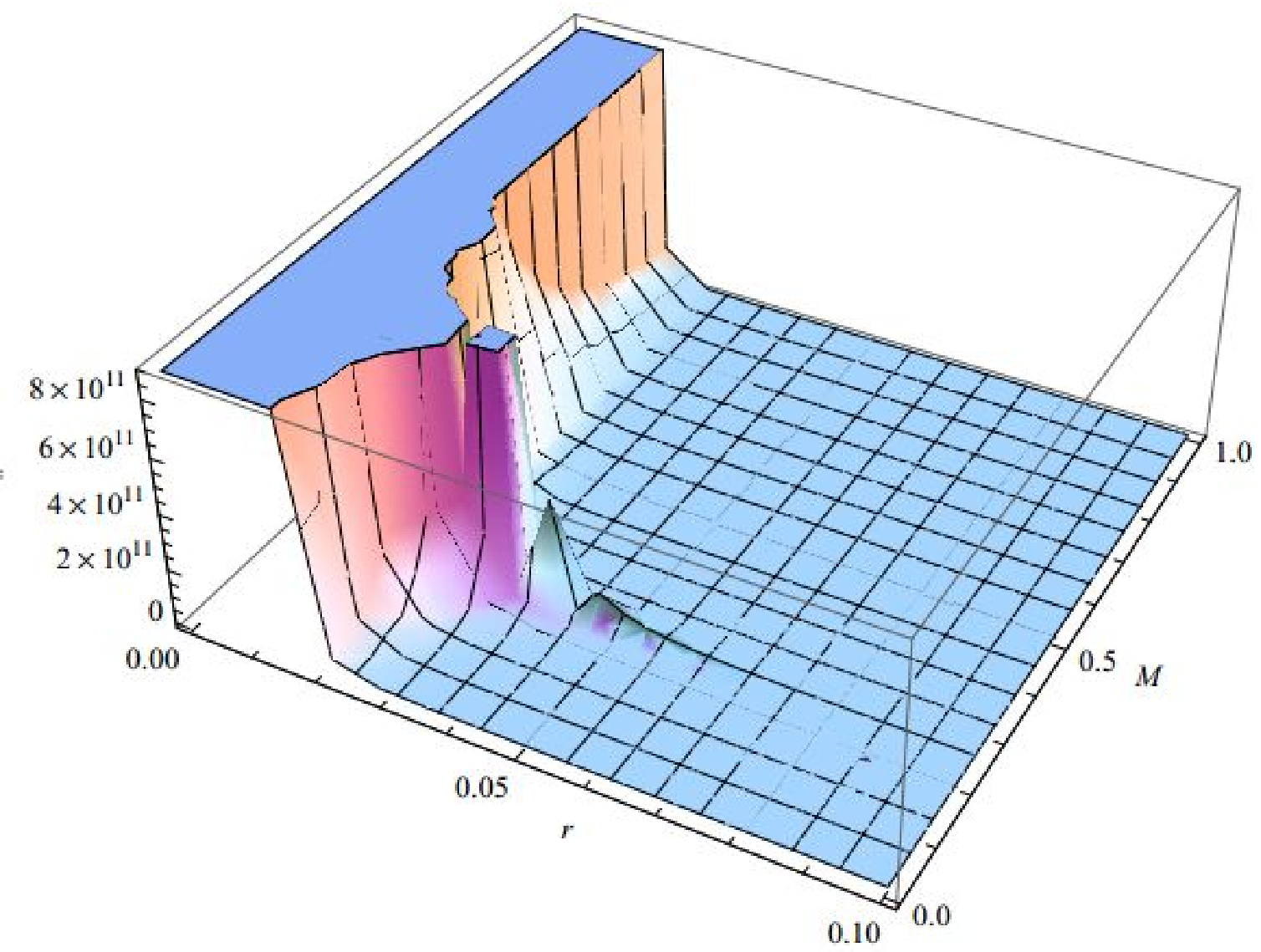}&
\includegraphics[width=8cm, height=8cm]{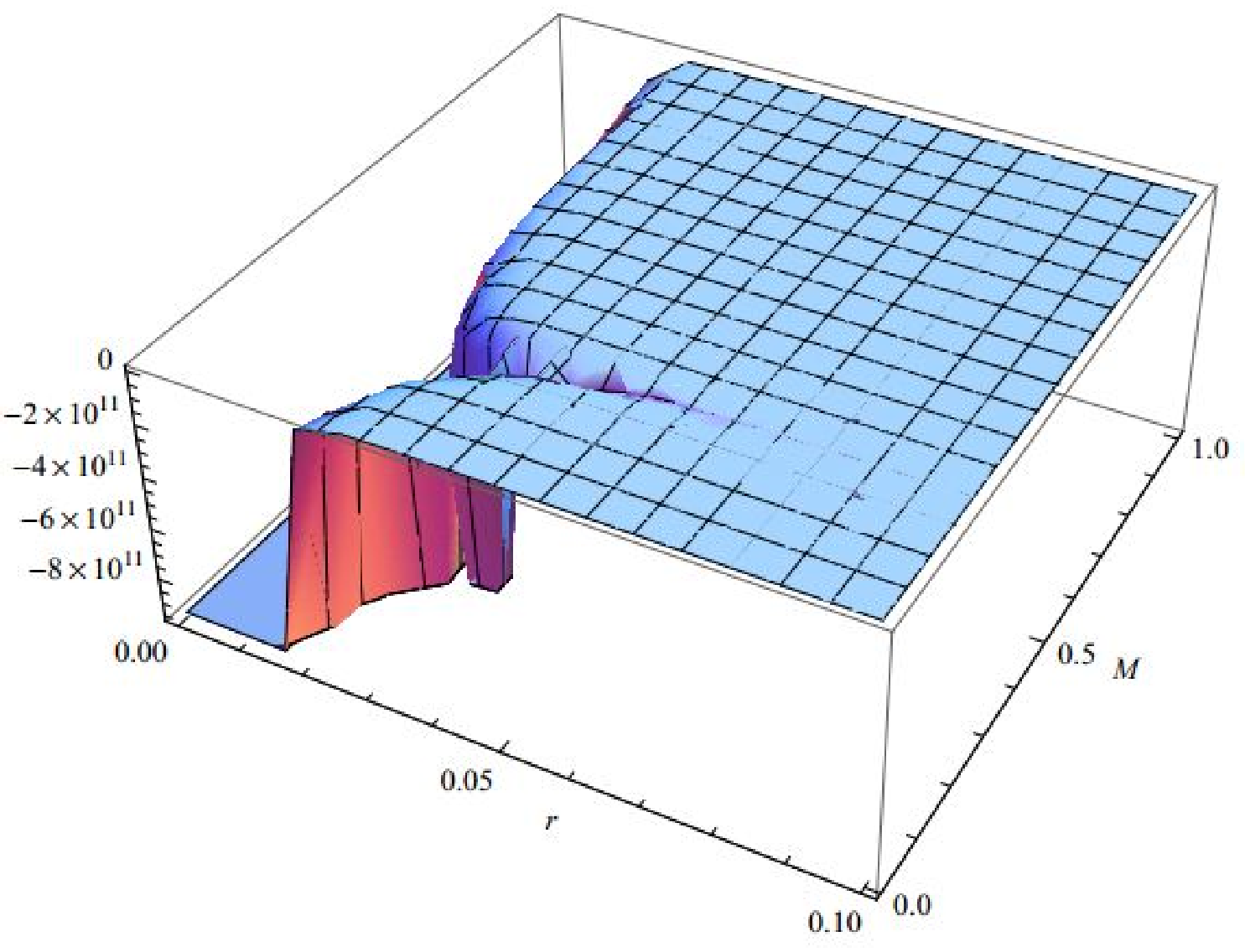}
\end{tabular}
 \caption{ The figures show the variation of the total energy $E$ with respect to the radius $r$ on $X$-axis and the mass $M$ on $Y$-axis 
  of the cosmic string for $\lambda=2$ ( figure at right ) and $\lambda=-2$ ( figure at left) respectively. 
  The graph is  plotted for $t=10^{12}$}
\label{fig2}
\end{figure}

\begin{figure}[h]
\centering
\includegraphics[width=8cm, height=8cm]{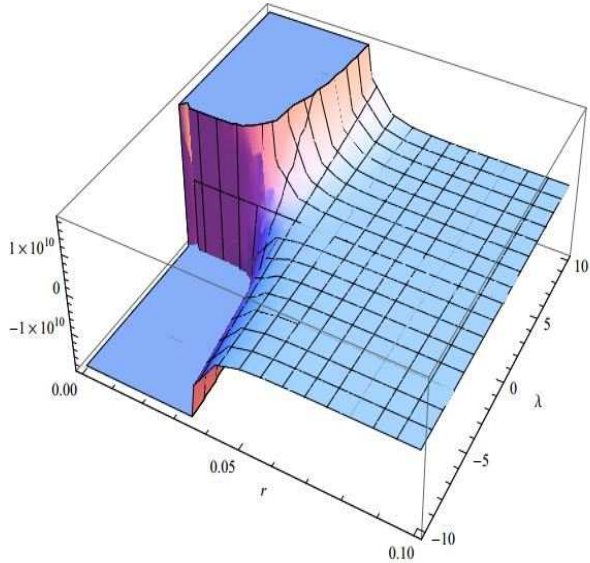}
  \caption{ The figure shows The figure shows the variation of the total energy $E$ with respect to the radius $r$ of the cosmic string on $X$-axis
  and the parameter $\lambda$ on $Y$-axis. 
  The graph is  plotted for $M=0.1$ and $t=10^{12}$.}
\end{figure}
\newpage
A conclusion that follows from these figures is that the cosmic string  total energy  $E$  per unit length in the z direction 
 is essentially non-zero for small radius of the cosmic string. 
 Moreover, this energy, which is not too sensitive\footnote{This may result from  the low  considered value of the  cosmological constant }
 to the variation of cosmic time, increases strongly with the increase of the parameter 
  $\lambda$ of the chosen $f(T)$ model.

\section{Conclusion} 
 In this paper we have obtained a general expression for the Laudau-Lifshitz energy-momentum complex in the realm
  of Teleparallel modified gravity, the  so  called $f(T )$ theory (in analogy to the $f (R)$ theories). 
  Such an expression has never appeared in the literature. The corresponding energy density 
has been evaluated for three plane symmetric metrics. For the first vacuum solution which has vanishing scalar torsion, the energy density
is well defined and can vanish for certain $f(T )$ models. The  second vacuum  solution also with vanishing scalar torsion 
is  characterized by a vanishing energy
density in Teleparallel theory as in  $f(T)$ theory. These results are totally different from those obtained in GR by using the same metrics.
The last vacuum metric has constant scalar torsion and contributes to a well defined  generalized energy density in $f(T )$ theory. 
An application has been made  for an important  Born-Infeld model satisfying weak energy condition.
\par 
In the second part of cosmological application  of Laudau-Lifshitz energy-momentum complex in this  this work,  we have evaluated the 
energy density  for a non-static cosmic string space-time. By considering  a quadratic $f(T)$ model, we have found  the energy distribution
of thin cosmic strings according to our plotting results. 
It is an energy per unit length in the $z$ direction which depends on the radius $r$ and the unitary mass $M$ of the cosmic string. This energy
increases considerably with the parameter of the  considered quadratic $f(T)$ model.





\begin{center}
 \rule{15cm}{1pt}
\end{center} 


\end{document}